\documentclass[11pt]{article}
\usepackage{hyperref}
\pdfoutput=1
\begin{document}
\title{Interfacial instabilities of spreading dielectric droplets subjected to ionic bombardment}
\author{Seyed Reza Mahmoudi$^1$, Sushant Anand$^1$, Kazimierz Adamiak$^2$,  
\\ G.S. Peter Castle$^2$, and Kripa K. Varanasi$^1$ \\
$^1$Massachusetts Institute of Technology, Cambridge, MA 02139, USA
\\$^2$University of Western Ontario, London, ON N6A 3K7, Canada} \maketitle
\begin{abstract}
This fluid dynamics video is an entry for the Gallery of Fluid Motion of the 66th Annual Meeting of the APS-DFD. This work reveals new types of interfacial instabilities that occur during forced-spreading of dielectric films subjected to ionic bombardment. We demonstrated that keeping the ionic bombardment strength constant, the appearance of instability patterns dramatically changes by varying viscosity of the bombarded droplet.  
\end{abstract}

\section{Introduction}

Recently, it was also demonstrated that the spreading of dielectric liquids accelerates over solid surfaces under ionic bombardment caused by corona discharge [1]. When a potential difference above the corona discharge threshold level is applied between a sharp emitter electrode and flat grounded substrate with a deposited oil droplet in air, the surrounding air molecules become locally ionized around the sharp tip. A cloud of accelerated ions drifts towards the air/liquid interface and creates an interfacial pressure. The interfacial electric pressure resulting from the surface charge deposition due to bombardment and the normal component of electric field cause radial spreading of dielectric liquid films [1]. Although the ionic bombardment initially forces spreading of the film, as we shown here, the spreading accompanies by new type of electrohydrodynamic (EHD) instabilities. EHD interfacial instabilities in fluid interfaces have been vigorously investigated in the presence of Laplacian field (in absence of space charge) produced by parallel plate electrodes. However, for non-zero space charge in the presence of ionic bombardment, the literature is rare and the mechanism of instabilities is poorly understood. In the presence of ionic bombardment, the instability mechanism fundamentally changes, giving rise to a new set of very less known class of instabilities [2]. The instability patterns were shown to be drastically changes with changing viscosity of liquid. For low viscous liquids subjected to ionic discharge, we observed cell-type instability patterns and length scales of patterns are in the order of spreading film thickness. For more viscous liquids, the patterns are like complex labyrinth patterns.

\section{Experiment Description}
In the present work, a 22g hypodermic needle was used to establish corona discharge over the deposited droplet on a grounded substrate. The applied voltage and corona current was, 17 kV and 35 $\mu$A, respectively. The substrate was ITO coated glass. The working fluids were silicone oils of different viscosities. We performed no special pretreatment of purification or degassing to eliminate the possible contaminations or impurities. The experiments performed in room condition with  no control on level of dust and other level of contamination.


\begin{thebibliography}{99}


\bibitem{mahmoudi}
Mahmoudi, S.R., Adamiak,K., Castle, G.S.P. Spreading of a dielectric droplet through an interfacial electric pressure. Proc. R. Soc. A. 2011 467, 3257.

\bibitem{mahmoudi}

Reyes, F. V., Garcia, F. J. Pattern imaging of primary and secondary electrohydrodynamic instabilities. J.  Fluid Mech. 2006 549, 61.



\end{thebibliography}
\end{document}